\documentstyle[12pt,epsfig]{article}
\newcommand{\beq}{\begin{equation}}
\newcommand{\eeq}{\end{equation}}
\newcommand{\bea}{\begin{eqnarray}}
\newcommand{\eea}{\end{eqnarray}}
\def\half{{\textstyle{1\over 2}}}
\def\quarter{{\textstyle{1\over 4}}}
\def\as{\alpha_s}

\def\dy{\Delta y}

\def\vpti{\vec{p}_{T1}}
\def\vptii{\vec{p}_{T2}}
\def\ptisq{p_{T1}^2}
\def\ptiisq{p_{T2}^2}

\def\d{{\rm d}}
\def\sigmahat{\hat{\sigma}}
\def\rsone{\sqrt s_1}
\def\rstwo{\sqrt s_2}

\def\Dzero{D$\emptyset$}

\def\GeV{{\rm GeV}}

\def\cG{{\cal G}}
\def\cC{{\cal C}}
\def\cK{{\cal K}}
\def\cF{{\cal F}}

\def\cR{{\cal R}}
\def\cO{{\cal O}}
\def\lapprox{\lower .7ex\hbox{$\;\stackrel{\textstyle <}{\sim}\;$}}
\def\gapprox{\lower .7ex\hbox{$\;\stackrel{\textstyle >}{\sim}\;$}}

\begin{document}
\titlepage
\begin{flushright}
{DTP/97/116}\\
{UR-1510}\\
{hep-ph/9801304}\\
{January 1998}\\
{Revised April 1998}
\end{flushright}
\begin{center}
\vspace*{2cm}
{\Large {\bf The Collision Energy Dependence of Dijet \\ [2mm]
Cross Sections as a Probe of BFKL Physics }} \\

\vspace*{1.5cm}
Lynne H.\ Orr$^a$ and W.\ J.\ Stirling$^{b}$ \\

\vspace*{0.5cm}
$^a \; $ {\it Department of Physics and Astronomy, University of Rochester,
Rochester, NY~14627-0171, USA}\\

$^b \; $ {\it Departments of Physics and  Mathematical Sciences, 
University of Durham,
Durham, DH1~3LE, UK}\\

\end{center}

\vspace*{4cm}
\begin{abstract}
The dependence of the subprocess cross section for dijet production
at fixed transverse momentum on the (large) rapidity separation $\Delta y$ 
of the dijets can be used to test for `BFKL physics', 
i.e.\ the presence of higher--order
$(\as \log\Delta y)^n$ contributions. Unfortunately 
in practice these subtle effects
are masked by the additional, stronger dependence arising from the parton 
distributions. We propose a simple ratio test using two different 
collider energies
in which the parton distribution dependence largely cancels.
 Such a ratio does 
distinguish between the `asymptotic' analytic BFKL  and 
lowest-order QCD predictions.  However, when subasymptotic effects 
from overall energy-momentum conservation are included, 
the BFKL predictions for present collider energies 
change qualitatively. In particular, the real gluon emission contributions to
the BFKL cross section are suppressed. The proposed ratio therefore provides
an interesting laboratory for studying the interplay of leading-logarithm
and kinematic effects.
\end{abstract}

\newpage

Quantum Chromodynamics has been very successful in describing jet 
production in high 
energy collider experiments.  Perturbative QCD at fixed order  in the 
strong coupling constant $\as$ provides 
sufficient predictive power for a wide variety of high energy phenomena.
However in some  regions of phase space, large logarithms can  
multiply the coupling,  
 spoiling the good behavior of fixed-order expansions.  In certain cases these
 large logarithms can be resummed, as in the 
Balitsky, Fadin, Kuraev and Lipatov (BFKL)  equation \cite{bfkl}, and 
the predictive power of the theory is then in principle restored.  
The latter statement should be subject to experimental test,
since the 
predictions of BFKL  are distinct from those of fixed-order QCD.

Being asymptotic, the regions
for which BFKL resummation is presumed to be necessary have until
recently been beyond the reach of experiments, and therefore the 
predictions of BFKL have been difficult to test.   The last few
years, however, have seen experiments at the HERA $ep$ collider 
and the Fermilab Tevatron that
approach the kinematic regimes relevant to BFKL.   At HERA the relevant 
regime is  small parton momentum fraction $x$, where BFKL predicts
a sharp rise in the structure function $F_2$ as $x$ decreases.  
Further discussion can be found in Refs.~\cite{jkreview,vddreview} and
references therein; 
here we are concerned with BFKL physics at hadron colliders.  

At the
Tevatron $\bar{p}p$ collider, and at hadron colliders in general, 
BFKL resummation applies to 
dijet production when the rapidity separation $\Delta$ of the two jets
is large, $\Delta \gg 1$, while fixed-order QCD should be adequate for
$\Delta = \cO(1)$.  BFKL predicts, among other things, a rise in the dijet 
subprocess cross
section $\hat{\sigma}$ as $\Delta$ increases \cite{muenav}, in contrast to
the $\hat{\sigma} \to \ $constant behavior expected at lowest order.  In practice 
such behavior can be difficult to observe because $\sigmahat$
 gets folded in with parton distribution functions (pdfs), which 
decrease with $\Delta$ much more rapidly than the subprocess 
cross section increases.  In particular, the BFKL rise with  $\Delta$
gets killed by kinematic constraints.  The challenge is to find 
measurable quantities
in dijet production that are insensitive to the pdfs, but that retain the
distinctive behavior characteristic of BFKL resummation.  
Some possibilities have been discussed in 
Refs.~\cite{ds1,wjs,ds2,ds3,schmidt,os}, 
including the azimuthal decorrelation of the two jets:  the multiple
emission of soft gluons between the leading jets predicted by BFKL leads
to a stronger decorrelation than does fixed-order QCD, and the prediction is 
relatively insensitive to the pdfs.  

Another possibility is to look for the increase in 
$\sigmahat$ with $\Delta$ by considering different collision energies
\cite{muenav}.
The idea is to choose $\Delta$'s that correspond to the same parton
momentum fractions at different energies so that the pdf dependence is
the same for both, thereby allowing the $\Delta$ dependence to be 
extracted.  
The \Dzero\ collaboration has in fact recently reported the results of
a preliminary study of dijet cross section ratios at fixed parton 
momentum fraction \cite{ANNA}. 

In this paper we investigate the dependence of the dijet
cross section on collision energies in leading-order QCD and in
the BFKL approach.  We define a cross section ratio
at two collider energies in which the pdf dependence largely cancels,
and in which the predictions of leading-order QCD and `naive' BFKL can be 
distinguished. (By naive BFKL, we mean the predictions obtained by
resumming leading-logarithm
$(\alpha_s\Delta)^n$ contributions arising from the 
emission of soft real and virtual 
gluon contributions  in the absence of overall kinematic constraints.)
However, one difficulty with the BFKL approach is that even 
for the largest $\Delta$ values which are accessible at present, subasymptotic
effects, for example from energy-momentum conservation, are likely 
to be important.
This has led to `improved' BFKL dijet cross section calculations, in which
some of these effects are taken into account. In this study we will use the
BFKL Monte Carlo approach developed in Ref.~\cite{os} (see also \cite{schmidt}).
We  find that at present collider energies 
the predictions of naive BFKL are indeed
substantially modified when subasymptotic kinematic effects are included.

Our calculations will therefore give us a sequence of predictions --- 
full leading-order 
QCD, asymptotic
leading-order QCD, naive BFKL, improved BFKL ---  to confront experiment.
In what follows we develop each of these approximations in turn.

We begin with the lowest-order QCD inclusive two-jet production cross
section as a function of the two jet rapidities $y_1, y_2$ and their 
common transverse momentum $p_T$.  For simplicity we will consider the
symmetric situation where the two rapidities are equal and opposite.
Setting $y_1  =-y_2 =\half\Delta$ gives for the differential cross section
\beq
\left. {\d\sigma\over \d y_1 \d y_2 \d p_T^2}\right\vert_{y_1 = -y_2 = 
\half\Delta} = 
{ 1 \over 256  \pi\, p_T^4 \,
 \cosh^4(\half\Delta) }
\sum_{a,b,c,d =q,g}\; x f_a(x,\mu^2)\;x f_b(x,\mu^2)\; 
\overline{\sum}\;| {\cal M}(ab\to cd)|^2 ,
\label{eq:fulllo}
\eeq
where $\overline{\sum}$ denotes the appropriate sums and averages over colors 
and spins and $f_{a,b}(x,\mu^2)$ are the parton densities. 
Here
\beq
x =  {2 p_T \over \sqrt{s}}\; \cosh(\half\Delta) \; ,
\eeq
and the subprocess matrix elements are functions only of the rapidity
difference $\Delta$.
In the limit of large $\Delta$ 
the 
$gg$, $qg$ and $qq$ subprocess matrix elements squared become equal,
up to overall color factors of $C_A^2$, $C_AC_F$ and $C_F^2$ respectively.
This defines the 
{\it effective subprocess approximation} to Eq.~(\ref{eq:fulllo}):
\beq
\left. {\d\sigma\over \d y_1 \d y_2 \d p_T^2}\right\vert_{y_1 = -y_2 = 
\half\Delta} \simeq 
{ [x \cG(x,\mu^2)]^2 \over 256  \pi\, p_T^4 \,
 \cosh^4(\half\Delta) }\ 
\overline{\sum}\;  | {\cal M}(gg\to gg)|^2 \;,
\eeq
where 
\beq
\cG = g + \frac{4}{9} \sum_q (q + \bar q) \; .
\eeq

In an experiment (and in the BFKL formalism to be discussed below),
we are interested in events with jets above some  transverse momentum  
 threshold $P_T$.
Integrating over the jet transverse momentum $p_T > P_T$ 
then gives
\beq
\left. {\d\sigma\over \d y_1 \d y_2 }\right\vert_{y_1 = -y_2 = 
\half\Delta} \simeq 
 { \overline{\sum}\;| {\cal M}(gg\to gg)|^2  \over
  256  \pi \, \cosh^4(\half\Delta) } \ {1\over P_T^2}\ 
 \int_1^{X^{-2}} \; {\d u^2\over u^4} \;
\left. \left[  x \cG(x,\mu^2) \right]^2\right\vert_{x =Xu} \; ,
\label{lowest}
\eeq
with $X = p_T({\rm min})/{p_T({\rm max})} = 2P_T\cosh(\half\Delta)/\sqrt{s}$. 
In the asymptotic limit where  $\Delta$ is large,
 the $\Delta$ dependence in 
the prefactor disappears so that
\beq
{ \overline{\sum}\;| {\cal M}(gg\to gg)|^2  \over
  256  \pi \, \cosh^4(\half\Delta) } \ \to \ \half \pi C_A^2   \as^2 \; ,
\label{amplim}
\eeq
and hence
\beq
\left. {\d\sigma\over \d y_1 \d y_2 }\right\vert_{y_1 = -y_2 = 
\half\Delta} \simeq 
 { \pi C_A^2 \as^2 \over 2  P_T^2}\ 
 \int_1^{X^{-2}} \; {\d u^2\over u^4} \;
\left. \left[  x \cG(x,\mu^2) \right]^2\right\vert_{x =Xu} \; 
\equiv \sigmahat_0\ \cF(X,\mu^2)\; 
\label{lowestasy}
\eeq
 where 
\beq 
\sigmahat_0 = { \pi C_A^2 \as^2 \over 2  P_T^2}
\label{defsbis}
\eeq 
and $\cF$ contains the integration over the parton distribution functions.
Equation (\ref{lowestasy}) defines the asymptotic 
QCD LO cross section to which we 
will compare the BFKL predictions below.

Fig.~1 shows the QCD LO dijet cross sections calculated using
({\it i})   the $2\to 2$ matrix elements (Eq.~(\ref{eq:fulllo})), 
({\it ii})  the effective subprocess approximation (Eq.~(\ref{lowest})),
 and
({\it iii}) the asymptotic form of the effective subprocess 
      approximation (Eq.~(\ref{lowestasy})),
for two $p \bar p$ collider energies $\sqrt{s} = 630,\; 1800\; \GeV$.
The jet $p_T$ threshold is $P_T = 20 \; \GeV/c$, the partons are
CTEQ4L with $\mu =P_T$ \cite{cteq4}, and $\as$ is evaluated in leading order 
also at scale  $\mu =P_T$ (the numerical value is 0.170). 
The effective subprocess approximation is
seen to reproduce the exact result to within a few per cent over
the entire $\Delta$ range.  The asymptotic form is approached at large
$\Delta$, as expected.   

At higher orders in QCD perturbation theory the subprocess cross section
receives large logarithmic corrections from soft gluon emission in
the rapidity interval between the two leading jets. For fixed coupling
$\as$ we have
\beq 
\sigmahat = \sigmahat_0
 \; \left[ 1 + \sum_{n\geq 1} a_n \; (\as \Delta)^n +
\ldots \ \right] \; ,
\label{defs}
\eeq 
with
$\sigmahat_0$  as defined above. 
The BFKL formalism to be discussed below
resums the logarithms and in the leading-logarithm, 
fixed-coupling
`naive' approximation gives a subprocess cross section which
is not constant but 
grows asymptotically
 with $\Delta$: $\sigmahat \sim \exp(\lambda\Delta) $ with $\lambda
= 4C_A\ln 2 \as/\pi \approx 0.5$. 

At fixed $\sqrt{s}$ and minimum transverse momentum 
$P_T$, {\it both } $\sigmahat$ and $\cF$ depend
on $\Delta$. In fact because of the shape of the parton distributions
the latter quantity decreases rapidly with increasing $\Delta$ and 
vanishes at the kinematic limit $\Delta  = 2 \cosh^{-1}(\sqrt{s}/2P_T)$,
as in Fig.~1. 
It is therefore difficult to observe the relatively slow rise
with $\Delta$ of $\sigmahat$ (see for example Fig.~8 of Ref.~\cite{wjs}).
The original idea of Ref.~\cite{muenav} was to increase the collider energy
$\sqrt{s}$ as $\Delta$ increases such that $X$ and therefore $\cF$ 
remains fixed. Any observed rise in the cross section could then only
arise from higher-order contributions to $\sigmahat$.

If two collider energies are available ($\rsone$ and $\rstwo$ say)
 one can make use of this idea by comparing the dijet
cross sections at two rapidity separations ($\Delta_1$ and $\Delta_2$)
{\it for which the asymptotic leading-order cross sections are equal}.
Specifically, if for a given $\Delta_1$ we define $\Delta_2$ such 
that
\beq
{ \cosh(\half\Delta_1)  \over  \cosh(\half\Delta_2) }
= { \rsone \over \rstwo } 
\label{deltadef}
\eeq
then the cross sections defined by Eq.~(\ref{lowestasy}) will
be equal, i.e.\ $\d\sigma(\rsone,\Delta_1) = \d\sigma(\rstwo,\Delta_2)$.
Using for $\sqrt{s_1}$
 and $\sqrt{s_2}$, respectively,
the two Tevatron collider energies $630$ and $1800\; \GeV$, 
we have the following values 
for $\Delta_1$ and $\Delta_2$:
\begin{center}                    
\begin{tabular}{|c|c|c|c|c|c|c|c|c|}  \hline                     
  $\Delta_1 (\sqrt{s_1}=630\ \GeV)$  &   0.0  &      1.0       
   &      2.0
 &  3.0  &   4.0  & 5.0 & 6.0 & 6.8 \\ \hline
 $\Delta_2 (\sqrt{s_2}=1800\ \GeV)$     & 3.42  &  3.68&  4.33
 &   5.19      
&  6.13      & 7.11 & 8.10 &  8.90  \\ \hline
\end{tabular}
\end{center}                                                                   
Note
that for large $\Delta_1$ Eq.\ (\ref{deltadef}) 
reduces to $\Delta_2 = \Delta_1 
+\ln(s_2/s_1) = \Delta_1 + 2.1$, as can be seen in the table.

The equality of the asymptotic cross sections at lowest order for the 
two $\Delta$'s leads us to  define a cross section ratio
\beq
R_{12} ={  \d\sigma(\rsone,\Delta_1) \over  \d\sigma(\rstwo,\Delta_2)}
\label{ratiodef}
\eeq
as a function of $\Delta_1$, with $\Delta_2$ given by Eq.~(\ref{deltadef}).
This ratio has the advantage of being directly accessible 
experimentally, since it depends on the rapidity difference
between the two `outside' jets, a quantity more easily measured
than  for example the parton momentum fractions themselves.
By construction, $R_{12} = 1$ in asymptotic LO QCD, but 
notice that the ratio is {\it not} equal to unity for the leading-order
cross section at subasymptotic rapidity separations 
calculated using exact matrix elements, or using the effective
subprocess approximation. This is because  the prefactor
multiplying the integral in for example Eq.~(\ref{lowest})
only becomes independent of $\Delta$ at large $\Delta$. 
This is illustrated in Fig.~2, which shows $R_{12}$ as a function
of $\Delta_1$ for $\rsone = 630\; \GeV$ and $\rstwo = 1800\; \GeV$, 
using the parameters of Fig.~1. The effective subprocess approximation
curve is very close to the exact curve and is not shown. The ratio
approaches unity at large $\Delta_1$, as expected.
Not surprisingly, the `exact matrix element' ratio is insensitive
to the choice of parton distributions. The dashed line shows the ratio
obtained using GRV94LO distributions \cite{grv}. The curves are almost
identical. Similarly, the scale choice dependence is also very weak.
For example, using the more natural `running' choice $\mu = p_T$ instead of
$\mu = P_T$ in the parton distributions and in $\as$ has a negligible
effect on the ratios.

The BFKL formalism resums the leading-logarithm\footnote{In fact, sub-leading 
logarithms ($\as^n\Delta^{n-1}$ etc.) can also be resummed 
in principle \cite{NLL}. In practice,
however, only the leading contributions are known at present in a form which is 
phenomenologically useful.}  $(\as\Delta)^n$ higher-order
corrections which arise from multiple emission of real and virtual gluons
in the rapidity interval between the two leading jets. The theoretical details
of how this is achieved can be found elsewhere (see for example
Ref.~\cite{qcdbook}) and will not be repeated here.

In fact the naive BFKL prediction, in which for example subasympotic effects from
overall energy momentum conservation are ignored,  has an  analytic 
representation which leads to a simple expression for $R_{12}$ in the
large $\Delta_1$ limit. 
If the coupling  $\as$ is assumed
to be constant, all kinematic constraints are ignored,
 and only the leading logarithms
are resummed, then the result is 
\beq 
\sigmahat = \sigmahat_0
 \; C_0\left(\frac{\as C_A}{\pi} \Delta \right) \; ,
\label{eq:bfkl0}
\eeq 
with 
\bea
C_0(t) &=& {1\over 2\pi}\; \int_{-\infty}^{+\infty} \; {\d z\over z^2
+ \quarter } \; e^{2 t \chi_0(z)}, \nonumber \\
  \chi_0(z) &=& {\rm Re}\,  [\, \psi(1) -\psi(\half+iz)\,  ] \; .
\label{coeffs}
\eea
The asymptotic (large $t$, equivalently large $\Delta$) behavior is
\beq
C_0(t)\; \sim \; {1\over \sqrt{\half \pi 7 \zeta(3) t }} \; e^{4 \log 2 t}\; .
\label{asympC0}
\eeq
It follows immediately that the naive BFKL prediction for the ratio
$R_{12}$ at large rapidity separations is 
\beq
R_{12} = { C_0\left(\frac{\as C_A}{\pi} \Delta_1 \right)
         \over C_0\left(\frac{\as C_A}{\pi} \Delta_2 \right) }\; .
\label{eq:asyratio}
\eeq
The prediction is shown in Fig.~2. Note that now $R_{12}$ is well below the 
QCD LO curves, and in fact 
at first sight this appears to be a
 primary test of the presence of higher-order `BFKL-like'
 corrections to the dijet cross section. The formalism is of course
 only supposed to be valid for large $\Delta_1$ (which implies
 large $\Delta_2$ also) and so we only show the predictions
 for $\Delta_1> 2$, which is roughly where the leading-order
 prediction begins to approach its asymptotic limit $R_{12} = 1$.
 The asymptotic large $\Delta_1$ 
 behavior of the ratio is readily obtained from 
Eqs.~(\ref{asympC0}) and (\ref{eq:asyratio}):
\beq
R_{12} \longrightarrow  \left({ s_1 \over s_2}\right)^{\lambda} \; , 
\quad \lambda = \frac{\as C_A}{\pi} 4\log 2 \; .
\label{eq:asyratio2}
\eeq
One interesting aspect of this result is that one is tempted to 
use the experimentally measurable quantity
$\cK = \log(R_{12})/\log(s_1/s_2)$ to determine the effective
 BFKL exponent $\lambda$, i.e.
\bea
\left. \cK\right\vert_{\rm{LO\; asymp.}} & = & 0 \; , \nonumber \\
\left. \cK\right\vert_{\rm{naive\; BFKL}} & \to & \frac{\as C_A}{\pi} 4\log 2 =
0.45\; ,
\eea
where the numerical value is obtained from using $\as$ evaluated
 at $\mu = P_T = 20\; \GeV$, as in Fig.~2. However it is clear from the figure
 that the asymptotic behavior is attained only very slowly. For this choice
 of parameters we expect $R_{12} \to 0.39$ as $\Delta_1 \to \infty$, but
 we see that the ratio has only decreased to
 $R_{12} = 0.45$ at the kinematic limit $\Delta_1 = 6.9 $.
Furthermore, as we shall see below, an improved BFKL calculation does 
not lead to such a simple prediction.

So far we have considered the predictions of naive BFKL, which can be 
written analytically but at the cost of making several assumptions which
do not hold in experimental situations.  In particular, the analytic 
solutions result from integrating over arbitrarily large values of the 
transverse momenta of emitted gluons, and the analytic phase space integrations
at the subprocess level prevents a proper incorporation of parton 
distributions.  In addition, $\as$ is assumed to be fixed.  The BFKL
approach can be improved and these assumptions avoided by recasting
the BFKL cross section as an event generator \cite{schmidt,os} using
an iterated solution to the BFKL equation.  Kinematic constraints and 
the running of $\as$ can then be included. The effect is to restrict the 
growth of $\sigmahat$ with increasing $\Delta$ compared to naive BFKL,
since the emission of relatively large transverse momentum gluons,
which give a positive contribution to the resummed cross section,
is suppressed.
Details of the calculation and application to the azimuthal decorrelation
can be found in \cite{os}.

Here we repeat the calculation of the cross section and ratios $R_{12}$
using the Monte Carlo of Ref.~\cite{os} to get an `improved' BFKL prediction.
One feature of the MC calculation that will be relevant to our results
is worth mentioning.  The MC solution to the BFKL equation is obtained by 
separating the contributions from real gluon emission into `resolved' and
`unresolved' contributions, the idea being that below a certain energy
scale $\mu_0$,
emitted real gluons are not detectable in practice.  The contribution from
unresolved real gluons is combined with the virtual gluon contribution to 
give an overall suppressing form factor.  The differential cross section
takes the form, for fixed $\alpha_s$,\footnote{We show the fixed $\alpha_s$
expression for simplicity; see \cite{os} for the 
corresponding running $\alpha_s$ expression.}
\beq
{d \hat{\sigma}_{gg} \over d^2 p_{T1}d^2 p_{T2} d \dy}
 = {\alpha_s^2  C_A^2 \over \ptisq \ptiisq }\;
   \left[ {\mu_0^2\over \ptisq}\right]^{\as C_A\Delta/\pi}\;
   \cR(\vpti, \vptii, \dy) \;  ,
\label{eq:sighatos}
\eeq
and $\cR$ can be written as a sum over resolved real gluon emissions:
\beq
 \cR(\vpti, \vptii, \Delta ) = 
  \sum_{n=0}^{\infty} \cR^{(n)}(\vpti, \vptii, \Delta )\;  ,
\label{eq:b7os}
\eeq
with $\cR^{(0)} = \delta( \vpti +  \vptii )$ and $\cR^{(n)}$ for $n>0$ given
in \cite{os}. 
The point is that even for $n=0$ the form factor 
$\left[ {\mu_0^2\over \ptisq}\right]^{\as C_A\Delta/\pi}$ appears in the 
cross section and represents a suppression of the probability that
no resolvable gluons will be emitted.  The inclusion of resolvable gluon
contributions ($n \geq 1$) counteracts this suppression and removes
the dependence on $\mu_0$.

Fig.~\ref{fig:sigmabfkl} shows the dijet cross section as a function of 
$\Delta$ 
for the naive BFKL and improved BFKL MC
cases at the two collision energies, with 
$P_T=20\ \GeV$.
Asymptotic QCD LO is also shown for reference.
There are several features worth noting here.  First, the naive BFKL 
cross section (dashed curve) is always largest, because it includes the 
analytic 
subprocess cross section given in  Eq.~(\ref{eq:bfkl0}), which allows emission
of any number of gluons with arbitrarily large energies.  The curve
falls off rather than increasing because $\hat{\sigma}$ is multiplied
by parton densities, but even those incorporate only  lowest order 
kinematics in this case.

When exact kinematics for entire events are included in both the 
subprocess cross section and the parton densities, as in the BFKL 
MC (solid curve), there is a dramatic suppression of 
the total cross section.\footnote{The running of $\alpha_s$,
which we include, 
also contributes to the suppression, but it has a much smaller
effect than kinematics.}  In fact the suppression is so strong that 
it drives the BFKL MC cross section {\it below} that for asymptotic QCD LO.
The reason is due to simple kinematics: 
the QCD LO cross 
section contains only two jets, but the BFKL MC cross section also includes
additional jets, each of which increases the subprocess center-of-mass energy
and elicits a corresponding price in parton densities.
In the naive BFKL calculation, the contribution to the subprocess energy
from additional jets is ignored and their net effect is to combine
with  the virtual gluons to increase the subprocess cross section.
Interestingly, the effect is compounded by the fact that the (naive) BFKL
increase of $\sigmahat$ only starts at $\cO(\as^2)$ in perturbation theory,
i.e. $C_0 = 1 + \cO(\as^2)$ in Eq.~(\ref{coeffs}). The  $\cO(\as^{1})$ 
correction is zero because the real and virtual gluon contributions
exactly cancel. However the effect of the pdfs is to suppress only
the real contributions, giving an overall negative correction at this order.

The ratio $R_{12}$ as calculated in the improved BFKL MC is shown in
Fig.~\ref{fig:ronetwobfkl}, again with the naive BFKL and asymptotic
QCD LO predictions.  Here the effects of kinematic suppression and the 
parton distributions are quite dramatic.  The BFKL MC curve does {\it not}
fall between the naive BFKL and QCD LO curves as one might expect.  
Instead, it is everywhere greater than or equal to the QCD LO curve,
and in particular, $R_{12}$ is everywhere greater than  1.

This behavior can be understood as follows. 
We can represent $R_{12}$ schematically as
\beq
R_{12} \sim   {  1 + \as\Delta_1 \cC_1 + \ldots           
        \over    1 + \as\Delta_2 \cC_2 + \ldots            } \; ,
\label{eq:Rschematic}
\eeq
where the $\cC_i$ are coefficients modified by the effects of the kinematic
suppression. By the arguments given above, we expect $\cC_i < 0$.
In the formal\footnote{We do not of course expect the BFKL formalism
to apply for small rapidity separations; here our aim is simply
to explain the behavior of the predictions of the BFKL MC calculation in
Fig.~\ref{fig:ronetwobfkl}.} limit $\Delta_1 \to 0$, the nonleading
terms in the numerator
vanish, the denominator is $< 1$ and hence $R_{12} > 1$.
On the other hand, in the limit $\Delta_1 \to \Delta_{\rm max}$, there
is no phase space available for {\it any} gluon emission, hence $\cC_i \to 0$
and $R_{12} \to 1$.  The observation of a `BFKL effect' would
require a region of phase space where $\Delta_i \gg 1$ but $\cC_i \simeq 1$.
Increasing the collider energy at fixed transverse momentum threshold
would eventually lead to this situation. In such a region we would expect
to see $R_{12} < 1$, as predicted by the naive BFKL calculation shown in 
Fig.~\ref{fig:ronetwobfkl}. Finally, why does the BFKL MC curve in
Fig.~\ref{fig:ronetwobfkl} {\it rise} again at large $\Delta_1$,
instead of approaching $1$ as simple kinematics would suggest?
The origin of this behavior arises in the form factor 
(see Eq.~(\ref{eq:sighatos}))
which suppresses the probability of no real (i.e.\ resolved) 
gluon emission in the rapidity interval between the dijets. As $\Delta$ 
increases
towards its maximum allowed value 
there is an effective upper (kinematic) limit on the transverse momentum
of each emitted gluon. This in turn leads to a suppressing  form factor with 
 the `artificial' parameter $\mu_0$ replaced by a physical parameter
 determined by kinematics and the pdfs. 
In other words, the all-orders leading-log contribution from virtual
gluon emission that is built into the overall form factor is 
not completely compensated by corresponding real emission because
the kinematics do not allow it.
 The kinematic suppression increases 
 with  $\Delta$, hence $\Delta_2 > \Delta_1$ implies $R_{12} > 1$.

The arguments above illustrate the general behavior of the cross section
ratio -- starting out at a value greater than 1, falling and then increasing
again at large $\Delta$ -- as predicted by the `improved'
BFKL Monte Carlo calculation.  The details, however -- how far the 
ratio falls (and in particular whether it
falls below unity as predicted in the naive calculation) and 
for how long -- depend on the specifics of the  experimental 
configuration, such as the collider energies and the transverse 
momentum threshold. In this study we have restricted ourselves to
parameter values which are currently accessible at the Tevatron. 
Clearly it would be desirable to have more than the two Tevatron
$p \bar p$ energies available for such a measurement.
In principle, the LHC at 14 TeV and RHIC in $pp$ mode at 500 GeV
would expand the range of  $\Delta$.  In practice, 
one is limited by $\Delta_{\rm max}$ at the lower energy.
Furthermore, comparing jet measurements at different machines with different
detectors can be challenging, and
ideally one would like to construct ratios of measurements made
at a single machine in a single detector, so that some systematic
uncertainties would cancel.   For example, running the LHC at
10 and 14~TeV would
make such a measurement possible, although ideally the two
energies should be more different. The essential point is that
higher collider energies appear to be necessary in order to prevent
the BFKL effects being swamped by kinematical constraints.

In summary, we have shown that  the effects of the  
increase in the dijet
subprocess cross section predicted in the naive BFKL approach can  in
principle be 
detected by measuring the ratio of 
cross sections at  energies and rapidities chosen such that 
the asymptotic QCD LO ratio is equal to 1.  Forming the ratio
minimizes the effects of the parton distribution functions.
However an improved BFKL calculation which includes the subasymptotic 
effects of the conservation of energy and momentum  gives a result 
for the ratio that is 
qualitatively different from that of naive BFKL.  In particular,
for dijet production at the two Tevatron energies,
the improved BFKL calculation gives either agreement with 
lowest-order QCD  or deviations  {\it opposite} 
 to those predicted by naive BFKL. 
If our arguments about the kinematic suppression of higher-order real gluon 
emission are correct, then it would appear that a fixed-order perturbation 
theory approach would be a more appropriate calculational framework
for the cross section ratio 
{\it at present collider energies}
than all-orders resummation.
Therefore, an important next step is to perform
a calculation of $R_{12}$ in exact next-to-leading-order QCD, to determine
whether it provides a satisfactory description of the data. 

Finally we emphasise that this study has concentrated on the cross section
ratio only. It may well be that other quantities are better suited
to revealing underlying BFKL effects. Indeed, we note that 
the azimuthal decorrelation of the dijet pair \cite{ds1,wjs}, 
as calculated in the naive BFKL approach,
is already non-zero at $\cO(\as)$ \cite{wjs}, and that its effects are not 
completely swamped by kinematics in an improved BFKL calculation \cite{os}.

\vspace{1.0cm}
\noindent {\bf Acknowledgements} \\

\noindent We acknowledge many useful discussions with members 
of the \Dzero\ collaboration,
and in particular with Anna Goussiou, whose ideas motivated this study.
LHO is grateful to the UK PPARC for a Visiting Fellowship.
This work was supported in part by the U.S. Department of Energy,
under grant DE-FG02-91ER40685 and by the U.S. National Science Foundation, under
grant PHY-9600155.


\newpage


\begin{figure}[t]
\begin{center}
\mbox{\epsfig{figure=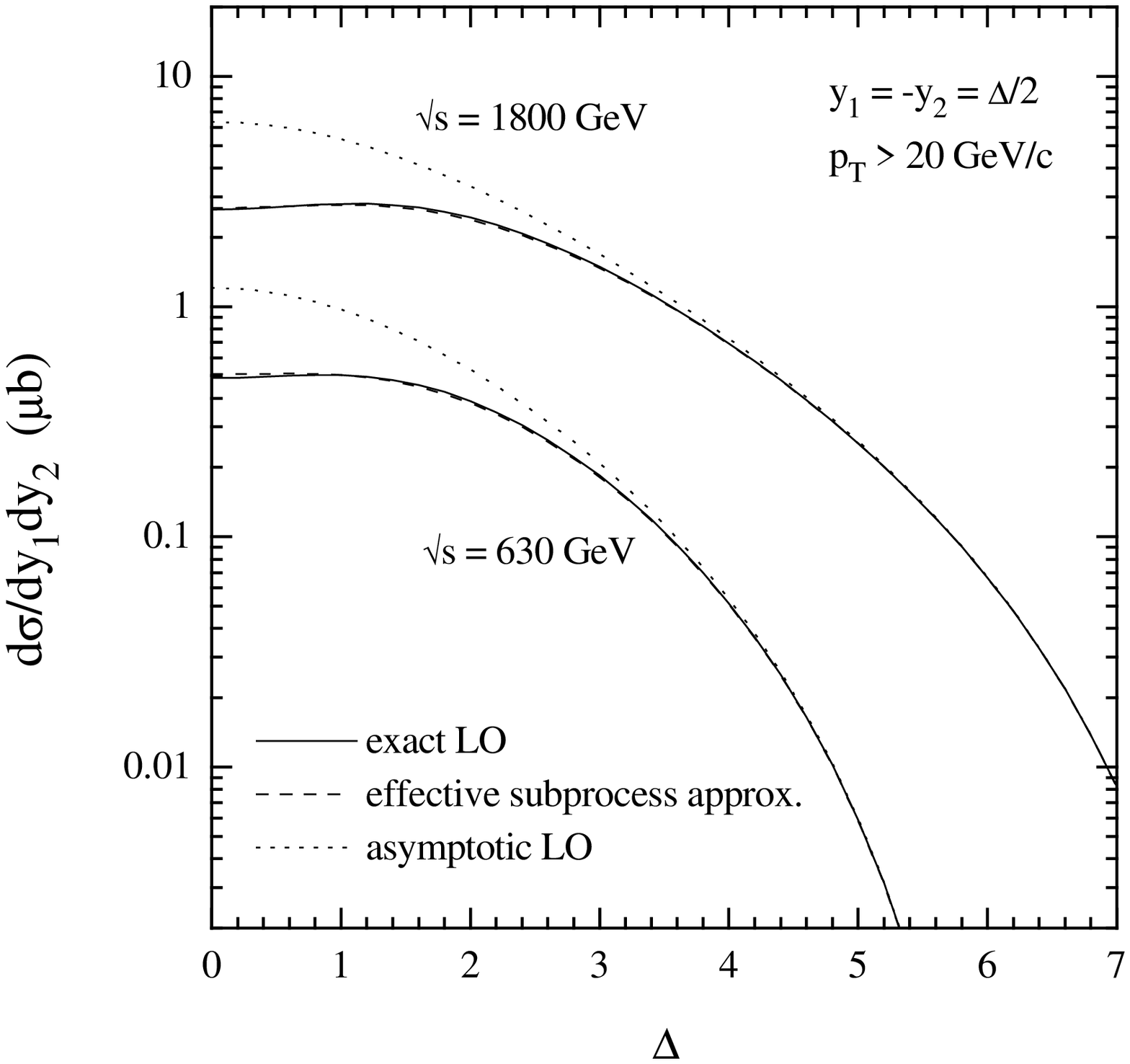,width=16.0cm}}
\caption[]{
The dependence of the leading-order ($2\to 2$) dijet cross sections  
on the dijet rapidity separation. The partons are the CTEQ4L set
 \protect\cite{cteq4}. The three curves at each collider energy
 use: (i) exact matrix elements (solid lines), (ii) the effective subprocess 
 approximation (dashed lines), and (iii) the asymptotic ($\Delta \gg 1$) form
 of the latter (dotted lines).
}
\label{fig:sigmalo}
\end{center}
\end{figure}


\begin{figure}[t]
\begin{center}
\mbox{\epsfig{figure=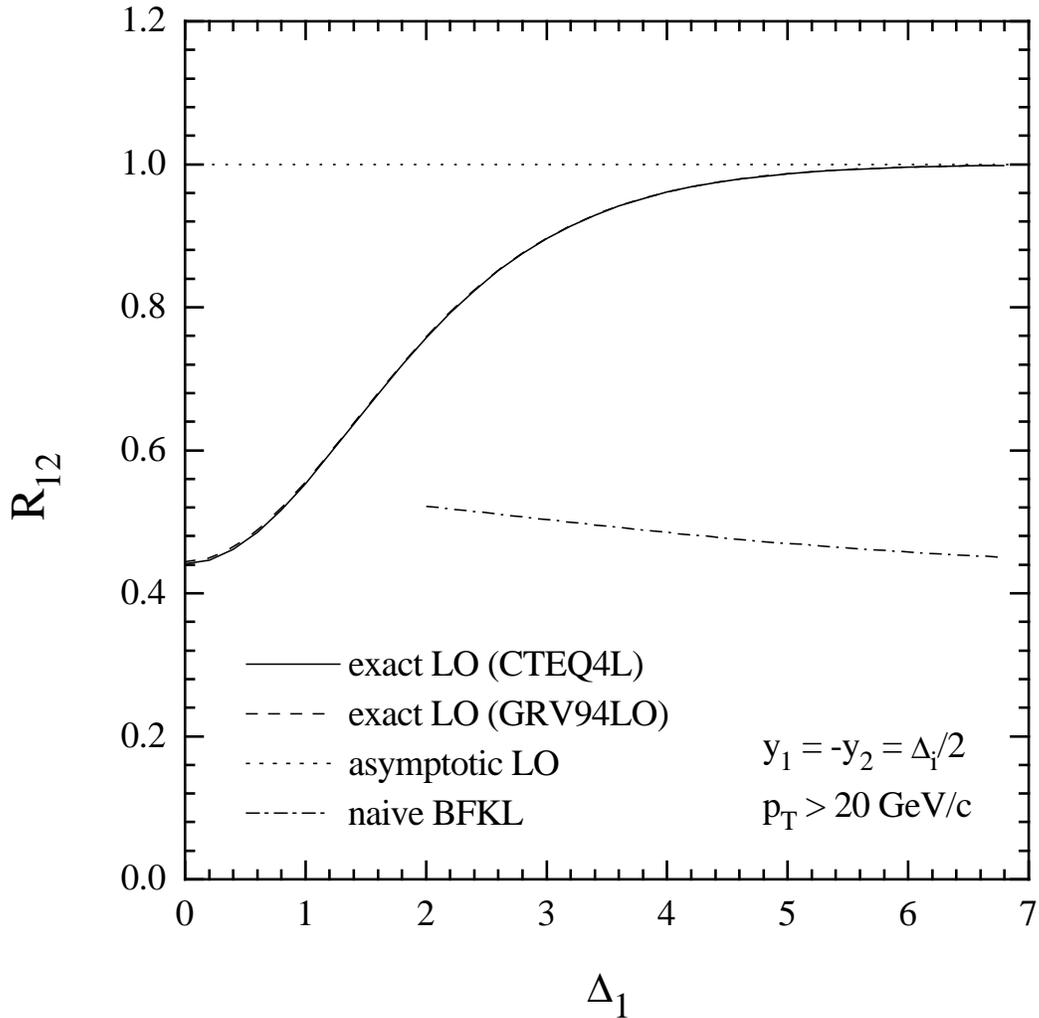,width=16.0cm}}
\caption[]{
The  ratio $R_{12}$ of the dijet cross sections at the two collider energies
$\rsone = 630\; \GeV$ and $\rstwo=1800 \; \GeV$, as defined in the text.
The curves are: (i) the exact leading-order   ($2\to 2$) predictions using
CTEQ4L (solid curve) \cite{cteq4} and GRV94LO partons (dashed curve)
\cite{grv}, both with $\mu =P_T = 20\; \GeV$,  
and (ii) the `naive BFKL' prediction (dot-dashed curve). Note that the
asymptotic leading-order prediction is $R_{12}=1$.
}
\label{fig:ronetwo}
\end{center}
\end{figure}

\begin{figure}[t]
\begin{center}
\mbox{\epsfig{figure=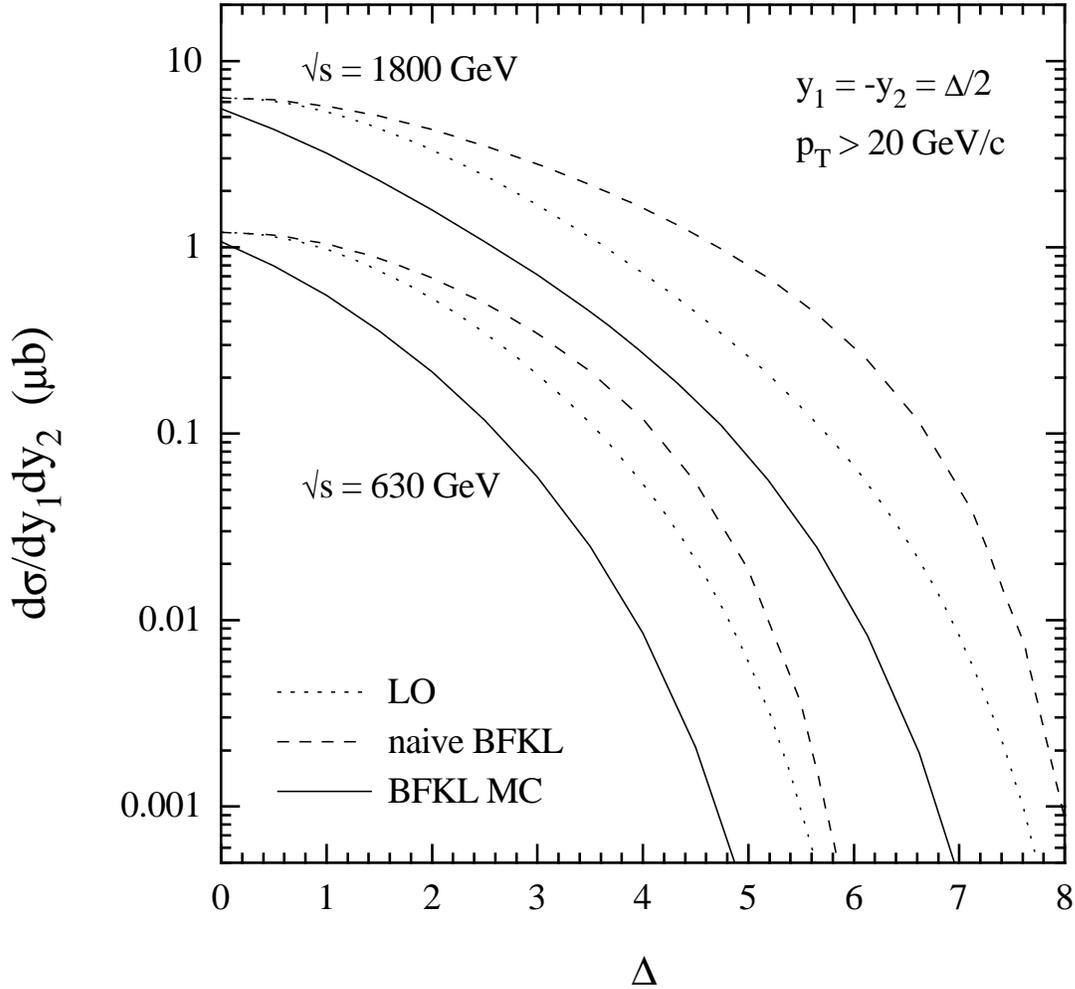,width=16.0cm}}
\caption[]{
The dependence of the BFKL and asymptotic QCD leading-order 
dijet cross sections  
on the dijet rapidity separation. The pdfs are the CTEQ4L set
 \protect\cite{cteq4}. The three curves at each collider energy
 use: (i) `improved' BFKL MC (solid lines), (ii) `naive' BFKL (dashed lines),
  and (iii) the asymptotic ($\Delta \gg 1$) form
 of QCD leading order (dotted lines).
}
\label{fig:sigmabfkl}
\end{center}
\end{figure}

\begin{figure}[t]
\begin{center}
\mbox{\epsfig{figure=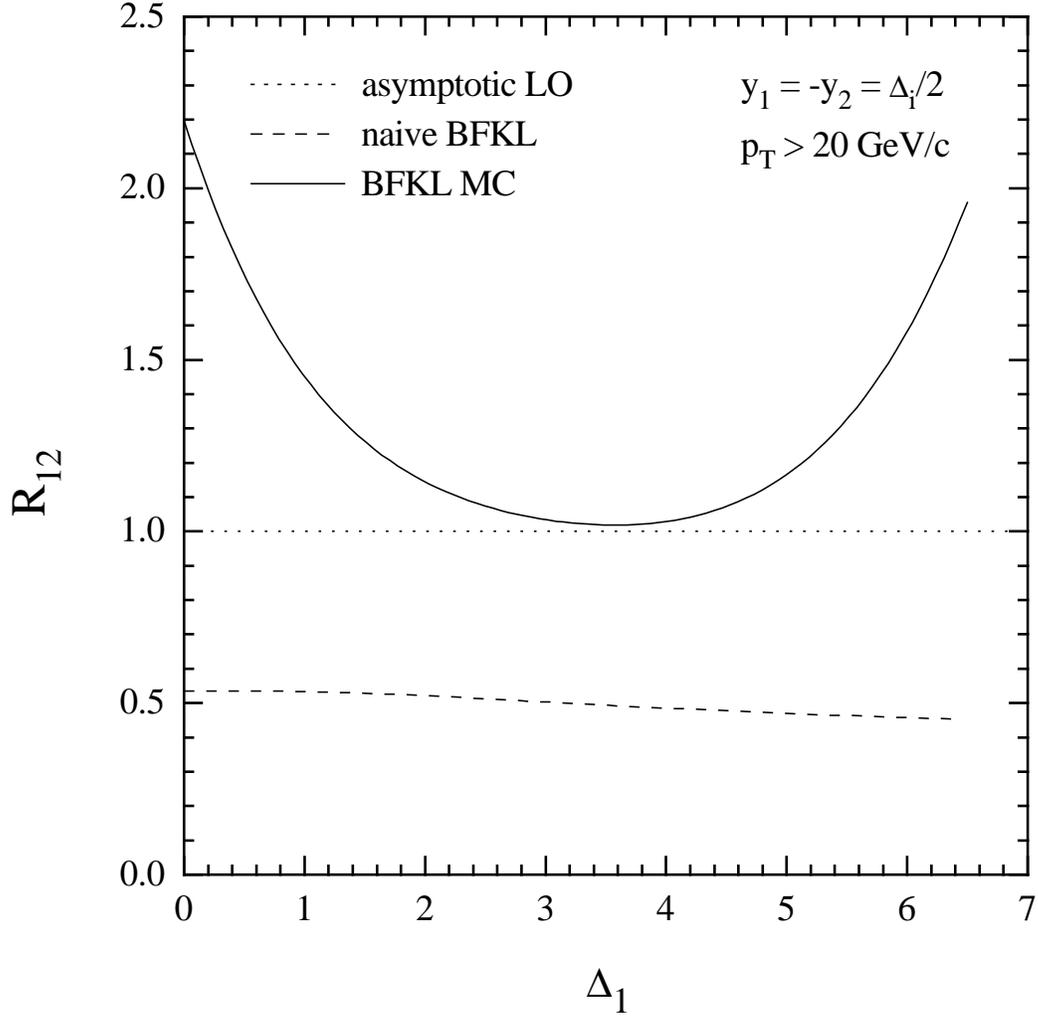,width=16.0cm}}
\caption[]{
The  ratio $R_{12}$ of the dijet cross sections at the two collider energies
$\rsone = 630\; \GeV$ and $\rstwo=1800 \; \GeV$, as defined in the text.
The curves are: (i) the `improved' BFKL MC predictions using
CTEQ4L \cite{cteq4}  pdfs (solid curve), 
 with $\mu =P_T = 20\; \GeV$,  
(ii) the `naive' BFKL prediction (dashed curve), and 
(iii)  the
asymptotic leading-order prediction (dotted curve)  $R_{12}=1$.
}
\label{fig:ronetwobfkl}
\end{center}
\end{figure}

\end{document}